\documentclass[
 reprint,
 superscriptaddress,
 amsmath,amssymb,
 aps,
prx,
floatfix,
]{revtex4-2}
\usepackage{graphicx}
\usepackage{dcolumn}
\usepackage{bm}
\usepackage{siunitx}
\usepackage{float}
\usepackage[T1]{fontenc}
\usepackage{mathptmx}
\usepackage{etoolbox}
\usepackage{amsfonts}
\usepackage{amsmath}
\usepackage{amssymb}
\usepackage{nicefrac}
\usepackage{enumitem}
\setlist[description]{font=\normalfont\itshape\space}
\usepackage{algorithm}[compatible]
\usepackage[noend]{algpseudocode} 
\usepackage[export]{adjustbox}
\usepackage{rotating}
\DeclareMathAlphabet{\mathcal}{OMS}{cmsy}{m}{n} 

\newcommand{\fd}[2]{{{\rm d}#1}/{{\rm d}#2}}
\renewcommand{\sd}[2]{{{\rm d}^2#1}/{{\rm d}#2^2}}
\usepackage{hyperref}

\begin{document}
\preprint{}
\title{Autonomous bootstrapping of quantum dot devices}

\author{Anton Zubchenko}
\affiliation{Center for Quantum Devices, Niels Bohr Institute, University of Copenhagen, Copenhagen 2100, Denmark}
\affiliation{QuTech and Kavli Institute of Nanoscience, Delft University of Technology, Delft, The Netherlands}
\author{Danielle Middlebrooks}
\affiliation{National Institute of Standards and Technology, Gaithersburg, MD 20899, USA}
\author{Torbj{\o}rn Rasmussen}
\affiliation{Center for Quantum Devices, Niels Bohr Institute, University of Copenhagen, Copenhagen 2100, Denmark}
\author{Lara Lausen}
\affiliation{Center for Quantum Devices, Niels Bohr Institute, University of Copenhagen, Copenhagen 2100, Denmark}
\author{Ferdinand Kuemmeth}
\affiliation{Center for Quantum Devices, Niels Bohr Institute, University of Copenhagen, Copenhagen 2100, Denmark}
\affiliation{Institute of Experimental and Applied Physics, University of Regensburg, 93040 Regensburg, Germany}
\affiliation{QDevil, Quantum Machines, 2750 Ballerup, Denmark}	
\author{Anasua Chatterjee}
\affiliation{Center for Quantum Devices, Niels Bohr Institute, University of Copenhagen, Copenhagen 2100, Denmark}
\affiliation{QuTech and Kavli Institute of Nanoscience, Delft University of Technology, Delft, The Netherlands}
\author{Justyna P. Zwolak}
\email{jpzwolak@nist.gov}
\affiliation{National Institute of Standards and Technology, Gaithersburg, MD 20899, USA}
\affiliation{Joint Center for Quantum Information and Computer Science,
University of Maryland, College Park, MD 20742, USA}
\affiliation{Department of Physics, University of Maryland, College Park, MD 20742, USA}

\date{\today}
\begin{abstract}
Semiconductor quantum dots (QDs) are a promising platform for multiple different qubit implementations, all of which are voltage controlled by programmable gate electrodes.
However, as the QD arrays grow in size and complexity, tuning procedures that can fully autonomously handle the increasing number of control parameters are becoming essential for enabling scalability. 
We propose a bootstrapping algorithm for initializing a depletion-mode QD device in preparation for subsequent phases of tuning. 
During bootstrapping, the QD device functionality is validated, all gates are characterized, and the QD charge sensor is made operational. 
We demonstrate the bootstrapping protocol in conjunction with a coarse-tuning module, showing that the combined algorithm can efficiently and reliably take a cooled-down QD device to a desired global-state configuration in under $8$ min with a success rate of $96~\%$. 
Finally, by following heuristic approaches to QD device initialization and combining the efficient ray-based measurement with the rapid radio-frequency reflectometry measurements, the proposed algorithm establishes a reference in terms of performance, reliability, and efficiency against which alternative algorithms can be benchmarked.
\end{abstract}

\maketitle
\section{Introduction}
Gate-defined quantum dots (QDs) in semiconductors present a promising avenue toward building high-fidelity and scalable quantum processors using existing industrial infrastructure~\cite{Loss98-QCD, Vandersypen19-QCS, Bavdaz22-QDC, Boter22-SWA, Ansaloni20-FFQ, Watson18-TQP, Xue21-CCC, Zwerver22-QMA, Noiri22-FUG, Philips22-UCS, Weinstein22-ULS, Takeda22-QES, Madzik22-PTT, Chatterjee21-AEC, Burkard21-SSQ}. 
However, due to the gate-defined and gate-controlled nature of spin qubits based on QDs, effective tuning within the large-dimensional gate-voltage landscape is becoming a key limitation to qubit array operation~\cite{Zwolak21-AAQ}. 
Previous work has made progress in areas such as topology and charge-state classification~\cite{Ziegler22-TRA, Zwolak20-AQD, Ziegler22-TAR}, navigation in higher-dimensional gate-voltage spaces using ray-based tuning~\cite{Zwolak21-RBI, Chatterjee21-AEC}, and optimization of readout and qubit fidelity~\cite{Berritta23-RTC, Hickie24-ALC}.  
However, most algorithms assume that 
previous tuning steps, such as the definition of the QD(s) channel(s), calibration of the charge sensor(s), or identifying regions of interest within the coarse-tuning regime, have been completed~\cite{Zwolak20-AQD, Zwolak21-RBI, Ziegler22-TAR}. 

In this work, we focus on the task of fully autonomous bootstrapping of the QD device, where ``bootstrapping'' is defined as a pretuning process that brings the device to an operational regime in which the coarse-tuning stage can be initiated~\cite{Zwolak20-AQD, Zwolak21-AAQ}. 
The proposed bootstrapping algorithm is designed to be initiated on a pristine QD device post-cooldown in the two-dimensional electron gas (2DEG) limit.
While previous work has proposed a machine-learning- (ML) based approach to QD device initialization~\cite{Moon20-ATQ}, our algorithm instead takes advantage of physics-based heuristics whenever possible.
The reason we choose to rely on domain knowledge rather than ML methods is that, at present, even the best-performing ML models are limited in their interpretability, making it challenging to determine how the QD device should be adjusted in the case of failure.
Moreover, low signal-to-noise ratio and other data imperfections can significantly affect the validity of ML predictions, causing unexpected failures of the tuning protocol.
Finally, automating bootstrapping following heuristic approaches allows us to establish a reference in terms of performance, reliability, and efficiency against which alternative, possibly more ML-heavy, algorithms can be benchmarked.
Thus, the procedural flow of the proposed bootstrapping algorithm closely follows the typical manual-tuning steps of a QD device.
Moreover, to improve the tuning efficiency and reliability, our algorithm utilizes a proximal charge sensor~\cite{Podd10-CSQ}, measured in radio-frequency (rf) reflectometry, instead of transport measurements used in previous work~\cite{Durrer19-ATQ}. 

\begin{figure*}
    \centering
    \includegraphics[width=\textwidth]{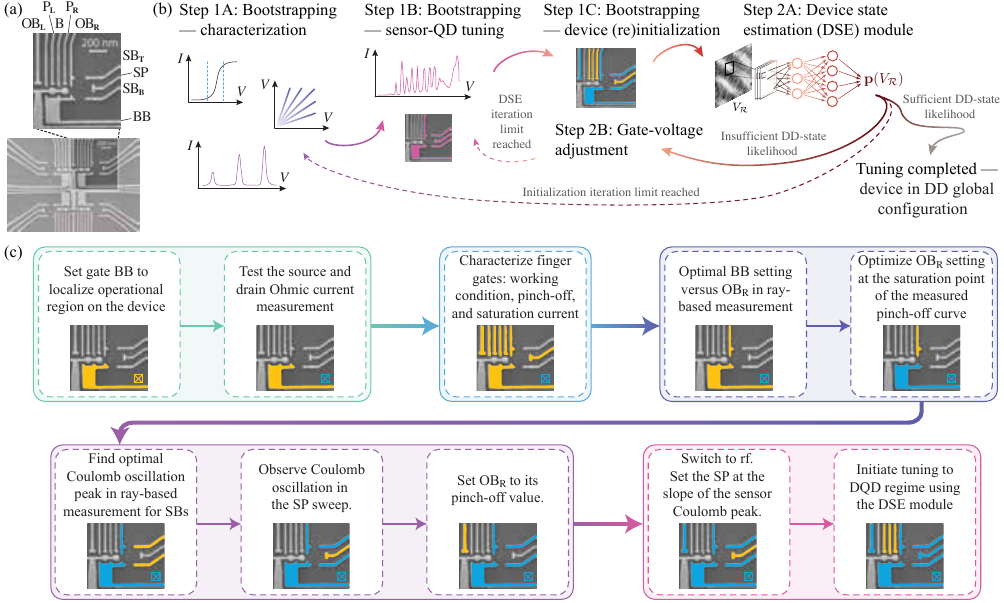}
    \caption{(a) A scanning electron micrograph (SEM) of a gate-defined, depletion-mode GaAs/Al$_{0.36}$Ga$_{0.64}$As device used to demonstrate the bootstrapping protocol.
    The gates are labeled as follows: P$_{\rm L}$, P$_{\rm R}$, and SP for the left, right, and sensor plunger, respectively; OB$_{\rm L}$, OB$_{\rm R}$, B, SB$_{\rm T}$, and SB$_{\rm B}$ for the outer left, outer right, middle, sensor top, and sensor bottom barrier, respectively; and BB for the backbone gate.
    (b) A conceptual overview of the tuning protocol.
    The bootstrapping process, depicted in steps 1A--1C, is followed by the device state estimation module (DSE; step 2A)~\cite{Zwolak20-AQD, Ziegler22-TRA}, and (optional) gate-voltage adjustment (step 2B).
    Steps 2 and 3 are interleaved with an optimization protocol and, if necessary, repeated characterization and reinitialization, to iteratively adjust voltages until the DSE module declares the device to be in a double-dot (DD) state. 
    (c) The procedural flow of the bootstrapping algorithm.
    The algorithm begins with the preparatory stage and setting the BB gate (teal box), followed by the device diagnostics and characterization of the finger gates stage (light blue box), defining of the sensor-QD- and qubit-QD-channels stage (dark blue box), a configuration of the outer barrier gates (purple box), and the sensor-tuning stage (pink box).
    }
    \label{fig:algorithm-flow}
\end{figure*}

The bootstrapping process starts after cooling the device down to a few millikelvins and involves determining the operational ranges, determining charging energies for each quantum dot (QD), and making the local charge-sensing system operational.  
In Fig.~\ref{fig:algorithm-flow}(b), we depict the flow of the tuning protocol reported in this work.
The procedural flow of the bootstrapping algorithm is shown in Fig.~\ref{fig:algorithm-flow}(c). 

\vspace{5pt}
\textit{Preparation.} 
The first, preparatory stage [teal box in Fig.~\ref{fig:algorithm-flow}(c)] consists of setting the backbone (BB) gate to localize the operational region, followed by testing and setting the Ohmic to a voltage that corresponds to the target current. 

\textit{Diagnostics and characterization.}
With the bias on the Ohmic set, the protocol proceeds with diagnostics and characterization of the device finger gates [light blue box in Fig.~\ref{fig:algorithm-flow}(c)].
At this stage, the algorithm tests whether all relevant gates are functional and collects pinch-off and saturation voltages where applicable.
Should any of the finger gates be flagged as non-functional, the bootstrapping process is terminated.

\textit{Defining QDs channels.}
Defining the sensor-QD and qubit-QD channel [dark blue box in Fig.~\ref{fig:algorithm-flow}(c)] begins from a pinched-off current using the BB and right outer barrier (OB$_{\rm R}$).
Then, the OB$_{\rm R}$ is fine tuned to the saturation point of the measured current.

\textit{Setting up charge sensor.}
The final stage of bootstrapping consists of two parts.
With the BB and OB$_{\rm R}$ set, the algorithm first sets the sensor barrier gates (SB$_{\rm T}$ and SB$_{\rm B}$) based on Coulomb oscillation in ray-based (rb) measurement and then the left outer barrier (OB$_{\rm L}$) in the qubit-QD channel to its pinch-off [purple box in Fig.~\ref{fig:algorithm-flow}(c)].
Finally, the measurement system switches to rf reflectometry, and the sensor plunger (SP) is calibrated [pink box in Fig.~\ref{fig:algorithm-flow}(c)].

Once the bootstrapping process is completed, the tuning algorithm initiates the coarse tuning~\cite{Zwolak20-AQD, Zwolak21-AAQ}.
The goal of coarse tuning is to find a range of gate voltages where the device is in a particular global configuration, i.e., a single-dot (SD) or double-dot (DD) state. 

We show that the proposed bootstrapping approach can be highly successful, with $86.5~\%$ of test runs successfully preparing the device for coarse tuning.
Given the rapid acquisition of data and its efficient analysis, our algorithm is capable of tuning the double-QD device within about $8$ min.
Moreover, using the parameters gathered during bootstrapping, we show a $96.6~\%$ success rate for coarse tuning, preparing the QD device for further calibration toward qubit operation.

Finally, we demonstrate the cross-platform applicability of the ML model used in the coarse-tuning module, which, thus far, has only been deployed on Si/Si$_x$Ge$_{1-x}$ QD devices with an overlapping gate architecture~\cite{Zwolak20-AQD, Ziegler22-TRA, Ziegler22-TAR, Ziegler23-AEC}. 
This work is, therefore, also a demonstration that an ML algorithm trained exclusively using synthetic data~\cite{Zwolak18-QLD} can be successfully used to assess the state across various device designs.

The remainder of the paper is organized as follows: In Sec.~\ref{sec:methods}, we give an overview of the design of the bootstrapping protocol.
The specifics of the algorithm and its configuration for all tests in this work are described in Secs.~\ref{ssec:boot_prep}--~\ref{ssec:boot_sens}.
The performance at the level of individual stages in bootstrapping is presented in Sec.~\ref{ssec:boot_perf}.
The overall performance of the tuning algorithm in reaching the target global configuration is discussed in Sec.~\ref{ssec:full_perf}.
We conclude with a discussion of the potential modifications to further improve the proposed bootstrapping technique in Sec.~\ref{sec:conclusion}.

\begin{figure*}
\centering
    \includegraphics[width=\textwidth]{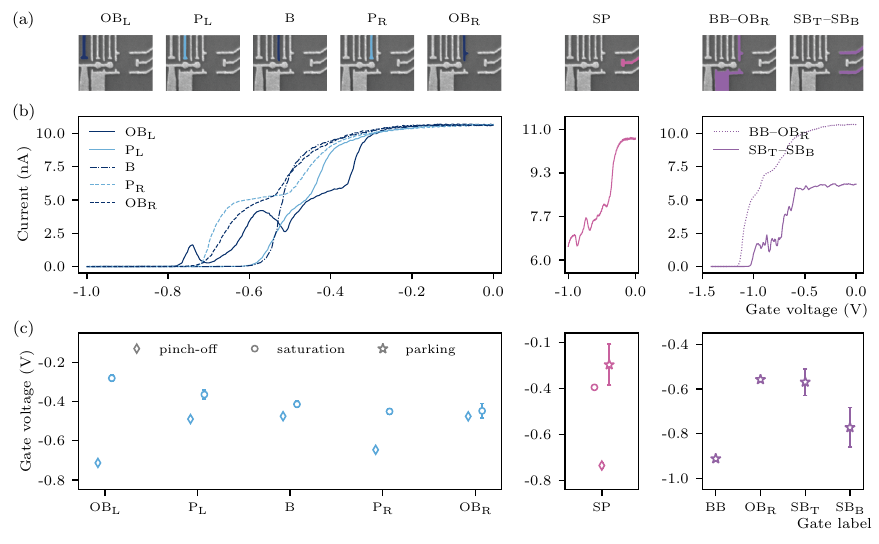}
    \caption{(a) SEMs highlighting finger gates characterized during the bootstrapping process. 
    (b) Sample voltage versus current plots for the finger gates depicted in panel (a). 
    The OB$_{\rm L}$ gate measures Coulomb oscillations due to the formation of a QD between OB$_{\rm L}$ and OB$_{\rm R}$.
    The BB, OB$_{\rm R}$, SB$_{\rm T}$, and SB$_{\rm B}$ gates are set as a result of ray-based (rb) measurement.
    SB$_{\rm T}$ versus SB$_{\rm B}$ measures Coulomb oscillations due to setting sensor-QD barriers.
    The BB versus OB$_{\rm R}$ rb measurement analyzes the current trace and finds the current saturation point: the gates are then parked at the voltage at which saturation is reached.
    The palette corresponds to the colors used to highlight gates in panel (a).
    (c) Consistency plots showing the mean values and standard deviations for the characteristics of interest, averaged over $N=89$ test bootstrapping runs. 
    For gates OB$_{\rm L}$, P$_{\rm L}$, B, P$_{\rm R}$, OB$_{\rm R}$, and SP, the pinch-offs and saturation values are collected.
    For SP, BB, OB$_{\rm R}$, SB$_{\rm T}$, and SB$_{\rm B}$, the parking-value statistics are collected.  
    For the SP, the parking value corresponds to the position on the most sensitive Coulomb peak position found with rf reflectometry.
    The final configuration for P$_{\rm L}$, B, and P$_{\rm R}$ are established during the coarse-tuning stage.
    The error bars indicate one standard deviation.
    However, their respective pinch-offs and saturation points are used to determine the starting position for the initialization of the DSE module.
}
    \label{fig:boot-stats}
\end{figure*}

\section{Methodology: Experimental setup}
\label{sec:methods}
Bootstrapping is initiated with a configuration file that defines the device architecture, identifies the quadrant that will be used in the experiment, defines the functionality of the gates (e.g., labeling gates as barriers or plungers), and specifies the safety ranges of the device. 
A scanning electron micrograph (SEM) of a gate-defined, depletion-mode GaAs/Al$_{0.36}$Ga$_{0.64}$As device used to demonstrate the bootstrapping protocol is shown in Fig.~\ref{fig:algorithm-flow}(a)~\cite{Fedele21-SOT}.
The two-dimensional electron gas (2DEG) is formed 57~\si{\nano\meter} below the surface.
The device, shown in Fig.~\ref{fig:algorithm-flow}(a), consists of four identical quadrants, each with tunable metallic electrodes for forming two QDs for hosting qubits and a larger sensor QD.
Barrier gates [OB$_{\rm L}$, B, OB$_{\rm R}$, SB$_{\rm T}$, and SB$_{\rm B}$ in Fig.~\ref{fig:algorithm-flow}(a) for the outer left, middle, outer right, sensor top, and sensor bottom barrier, respectively] confine each of the QDs, while plunger gates [P$_{\rm L}$, P$_{\rm R}$, and SP in Fig.~\ref{fig:algorithm-flow}(a) for the left, right, and sensor plunger, respectively] control their chemical potentials.
The backbone gate and an appropriate separator gate [BB and OB$_{\rm R}$ in Fig.~\ref{fig:algorithm-flow}(a)] are used to isolate the quadrant of interest.
The separator gate is also used as a barrier for one of the qubit QDs as the designated gate was dysfunctional.
Each quadrant has two Ohmic contacts to the 2DEG which are also connected to an off-chip resonator.
The Ohmics are used to measure the current through the 2DEG. 
The resonator and sensor QD form an RLC circuit for rf-reflectometry measurements.

The experiment is conducted in the dilution fridge at roughly $30$~\si{\milli\kelvin}.
The Quantum Machines QDAC-II is used to set the bias on the source Ohmic, as well as to perform voltage sweeps of the gates. 
The current measurements are acquired and converted into voltage readings using an Ithaco 1211 current preamplifier before being fed into a Keysight 34465A 6.5 digital multimeter.
Finally, the rf-reflectometry measurements for the DSE-based tuning protocol are acquired using a Quantum Machines OPX tuned to $273.8$~\si{\mega\hertz}, the resonant frequency of the selected quadrant.

\subsection{Bootstrapping: Preparation}
\label{ssec:boot_prep}
Once the device is cooled down, the preparatory stage begins [teal box in Fig.~\ref{fig:algorithm-flow}(c)]. 
The localization of the operational quadrant is achieved by setting the BB gate to the most negative safe voltage of the gate.
Then, with all gates set to zero, the bias voltage is swept to test for Ohmic behavior.
If the expected linear relationship between the voltage bias and the measured current is observed, the Ohmic is biased to the voltage where a small positive current of about $10$~\si{\nano\ampere} can be observed. 
This amperage was determined during the manual prescreening to be both safe for the device, and sufficiently high for measurements using the experimental setup described in Sec.~\ref{sec:methods}.

\subsection{Bootstrapping: Diagnostics and characterization}
\label{ssec:boot_char}
With the bias on the source Ohmic set, the algorithm proceeds to assess the functionality of the finger gates in the quadrant, as shown in the light blue box in Fig.~\ref{fig:algorithm-flow}(c).
This is achieved by applying a voltage to each gate, one at a time, and measuring the current through the device.
Ideally, the voltage versus current signal should have a sigmoidal shape.
However, the device imperfections may lead to curves that more or less significantly violate this assumption, as can be seen in Fig.~\ref{fig:boot-stats}(b).

The analysis of the current versus voltage signal serves several purposes.
First, it is a diagnostic tool that verifies whether gates are functional and the device is suitable for tuning.
Finger gates are classified as working if they produce a current versus voltage signal with a sufficiently wide distribution, as quantified by the median absolute deviation (MAD). 
The MAD is compared to the threshold value that is expected to be higher than the experimental setup noise levels determined from the device prescreening to be less than $1~\%$ of the target current. 

Second, all the relevant gate characteristics, such as pinch-off, saturation, and parking values, are determined as part of the bootstrapping process. 
To determine the pinch-off and saturation values, we consider the first ($\fd{I}{V}$) and the second ($\sd{I}{V}$) derivative of the current versus voltage signal, respectively.
Prior to analysis, the gradient of the signal is first smoothed with a Gaussian filter and then fitted with an inverse cosine hyperbolic function of the form
\begin{equation}
    f(x) = \frac{a}{\cosh^2[b(x-x_o)]} + y_o,
\end{equation}
where $a$, $b$, $x_o$, and $y_0$ are fit parameters used for a \textit{post-hoc} analysis~\cite{Sohn13}.
The location of the minimum-voltage prominent peak in the $\fd{I}{V}$ curve defines the pinch-off value. 
The saturation point value is defined at the location of the maximum-voltage negative prominent peak of the $\sd{I}{V}$.
The prominence assessment is used to identify stand-out features. 
For finding the pinch-off and saturation values, calculated prominence values are used to filter the peaks arising from the experimental setup noise. 
Additionally, prominence analysis is used for finding and narrowing down the candidate list of optimal Coulomb oscillation peaks during the sensor-QD tuning that uses a threshold value corresponding to $1~\%$ of the target current (see Sec.~\ref{ssec:boot_sens}).

\subsection{Bootstrapping: Defining QDs channels}
\label{ssec:boot_QD_chan}
With all gates characterized, the bootstrapping algorithm proceeds to define the QDs channels [dark blue box in Fig.~\ref{fig:algorithm-flow}(c)]. 
The goal of this step is to adjust the BB and OB$_{\rm R}$ gates to split the QD device quadrant into the sensor-QD and qubit-QD channels. 

To find the combined voltages for BB and OB$_{\rm R}$ that pinch off current, the algorithm utilizes the rb measurement instead of the conventional 2D scans. 
The rb approach involves measuring a collection of one-dimensional (1D) voltage sweeps initiated from a common starting point $(V_{\rm BB}, V_{\rm OB_{R}}) = (0, 0)$ at a tilt from the central measurement axis, defined as the straight line to  $(V_{\rm BB}, V_{\rm OB_{R}}) = (-1, -1)$.
The tilt angles, set at seven evenly spaced orientations in a range between $0.2\pi$ and $0.3\pi$, are chosen to increase the likelihood of observing the current pinch-off.
The algorithm picks the angle for which the pinch-off voltage is the lowest, decomposes the optimal pinch-off voltage into the BB and OB$_{\rm R}$ voltage components, and then sets the BB gate based on this decomposition.

With BB calibrated, a full trace of the OB$_{\rm R}$ gate is measured and OB$_{\rm R}$ is parked at the position of the peak in the $\fd{I}{V}$ with the highest current for sustaining transport through the QDs.

\subsection{Bootstrapping: Setting up charge sensor}
\label{ssec:boot_sens}
Once the current channels are established, the protocol proceeds with tuning the outer barrier gates for both the sensor and qubit QDs [purple blue box in Fig.~\ref{fig:algorithm-flow}(c)].
With the BB and OB$_{\rm R}$ gates calibrated and the SP gate parked at 0 V, rb measurement is conducted for SB$_{\rm T}$ versus SB$_{\rm B}$ in a similar fashion as for BB versus OB$_{\rm R}$.

The goal of this stage is the detection of the Coulomb blockade induced by the formation of the sensor QD, manifested through oscillations in the current sweep measured on the SP gate.
If oscillations are absent in all rays, the protocol is programmed to abort the run and restart. 
For all rays where the oscillations are detected, all peaks are analyzed to find one that is most likely to lead to a high sensitivity of the sensor.

All detected peaks are first filtered to remove overlapping peaks from further analysis. 
Then, the remaining peaks are ordered according to a score function inspired by Ref.~\cite{Baart16-CAT}:
\begin{equation}\label{eq:2}
    \mathcal{S} = h\,\frac{2}{1+hw/hw_0},
\end{equation}
where $h$ is the peak height, $hw$ is the full width at half maximum of the peak, and $hw_0 = 10$~\si{\milli\volt} is an expected full width at half maximum for the device type used to test the bootstrapping algorithm~\cite{Baart16-CAT}. 
The $h$ and $hw$ are determined based on the prominence analysis described in Sec.~\ref{ssec:boot_char}. 

Once the most promising peak is determined, SP is put back into the open position, and the OB$_{\rm L}$ is swept to localize qubit QDs.
Again, if oscillations are observed, the qubit QD has been successfully confined and, thus, it is possible to split it with the B gate to form DD. 
We then proceed to turn on the rf-reflectometry setup for charge sensing. 
To achieve the best sensitivity, the plunger needs to be parked at the largest slope of the signal reflected off the sensor QD. 
In this way, small shifts in the DD capacitances due to electron transitions can be observed.  

\subsection{Coarse tuning}
\label{ssec:coarse_tun}
Once the bootstrapping module is completed, the coarse-tuning stage begins~\cite{Zwolak20-AQD, Zwolak21-AAQ}.
The finger-gate characteristics from bootstrapping are used to define the operational regime by limiting the range over which gates P$_{\rm L}$, B, and P$_{\rm R}$ can be adjusted. 
The DSE-based tuning module is then initiated at voltages of the P$_{\rm L}$, B, and P$_{\rm R}$ gates randomly selected from a uniform distribution between $V_{\rm saturation} + \gamma$ and $V_{\rm saturation} - \gamma$ for each respective gate, where $\gamma=V_{\rm saturation}-V_{\rm pinch-off}$.
Proper scaling of the measurement scans is crucial for meaningful network analysis: scans that are too small may not contain enough features necessary for state classification, while scans that are too large may result in probability vectors that are not useful in the navigation phase.
The size of the scans is determined based on the charging energies extracted during the diagnostics and characterization phase. 
The scan size is set to match roughly $1.5$ times the initial charging energies of individual QDs in the qubit-QDs region of the device, as suggested in Ref.~\cite{Ziegler23-AEC}. 

The navigation algorithm within the DSE-based module is driven by the distance between the state vector describing the target state ${\rm\bf{p}}_{\rm target}$ and the state vector ${\rm\bf{p}}(V_\mathcal{R})$ predicted for current measurement.
The state vector quantifying the current state of the device is defined as 
\begin{equation}
    \bm{p}(V_\mathcal{R})=[p_{\rm ND},\,p_{{\rm SD}_{\rm R}},\,p_{{\rm SD}_{\rm C}},\,p_{{\rm SD}_{\rm L}},\,p_{\rm DD}],
\end{equation} 
with ND indicating no QDs formed, SD$_{\rm R}$, SD$_{\rm C}$, and SD$_{\rm L}$ denoting the right, central, and left single QD, respectively, and DD denoting the double-QD state.

The distance between the two state vectors is defined as:
\begin{equation}\label{eq:fit-func}
    \delta[{\rm\bf{p}}_{\rm target},{\rm\bf{p}}(V_\mathcal{R})] = 
    \left\lVert{\rm\bf{p}}_{\rm target} - {\rm\bf{p}}(V_\mathcal{R})\right\lVert_2 + \epsilon(V_\mathcal{R}), 
\end{equation}
where $\left\lVert\cdot\right\lVert_2$ is the Euclidean norm and $\epsilon(\cdot)$ is a penalty function for tuning to larger plunger voltages~\cite{Zwolak20-AQD, Zwolak21-RBI}.
The penalty is determined dynamically as a function of the distance to the middle of the operational regime defined by the bootstrapping.

Depending on the outcome of the DSE module, the algorithm can take one of three actions.
If the distance between ${\rm\bf{p}}_{\rm target}$ and ${\rm\bf{p}}(V_\mathcal{R})$ falls below a predefined threshold $\delta_{\rm tr}$, the tuning process is terminated and the device is declared to be in a DD global configuration. 
The choice of $\delta_{\rm tr}$ translates directly to how close in the state space the final point is to the DD region.
By default, $\delta_{\rm tr}=0.3$, which means that the estimated likelihood of the DD regime within the measurement is at least $85~\%$ DD state.

If the distance between ${\rm\bf{p}}_{\rm target}$ and ${\rm\bf{p}}(V_\mathcal{R})$ surpasses the predefined threshold, i.e., $\delta[{\rm\bf{p}}_{\rm target},{\rm\bf{p}}(V_\mathcal{R})] >\delta_{\rm tr}$, and the DSE module iteration limit has not been reached, the gate voltages for the plunger gates are adjusted [step 2B in Fig.~\ref{fig:algorithm-flow}(b)].
The navigation is driven by the action-based algorithm proposed in Ref.~\cite{Ziegler22-TAR}.

This process is repeated until either the DD state is declared or the DSE module iteration limit is reached.
Reaching the DSE module iteration limit triggers recalibration of the sensor QD, which is followed by reinitialization of the DSE module.
By default, the iteration limit is set to $10$.
The reinitialization involves finding new starting values for P$_{\rm L}$, B, and P$_{\rm R}$.
This is done by resampling the uniform distributions around saturation points for all three gates.

Finally, if $\delta[{\rm\bf{p}}_{\rm target},{\rm\bf{p}}(V_\mathcal{R})] >\delta_{\rm tr}$, the DSE module iteration limit has been reached, and the reinitialization iteration limit has been reached, the bootstrapping is restarted.
The autotuning is considered successful if the optimizer converges to a voltage range that gives the expected DD configuration when evaluated by human experts. 

\begin{figure*}
    \includegraphics[width=\textwidth ]{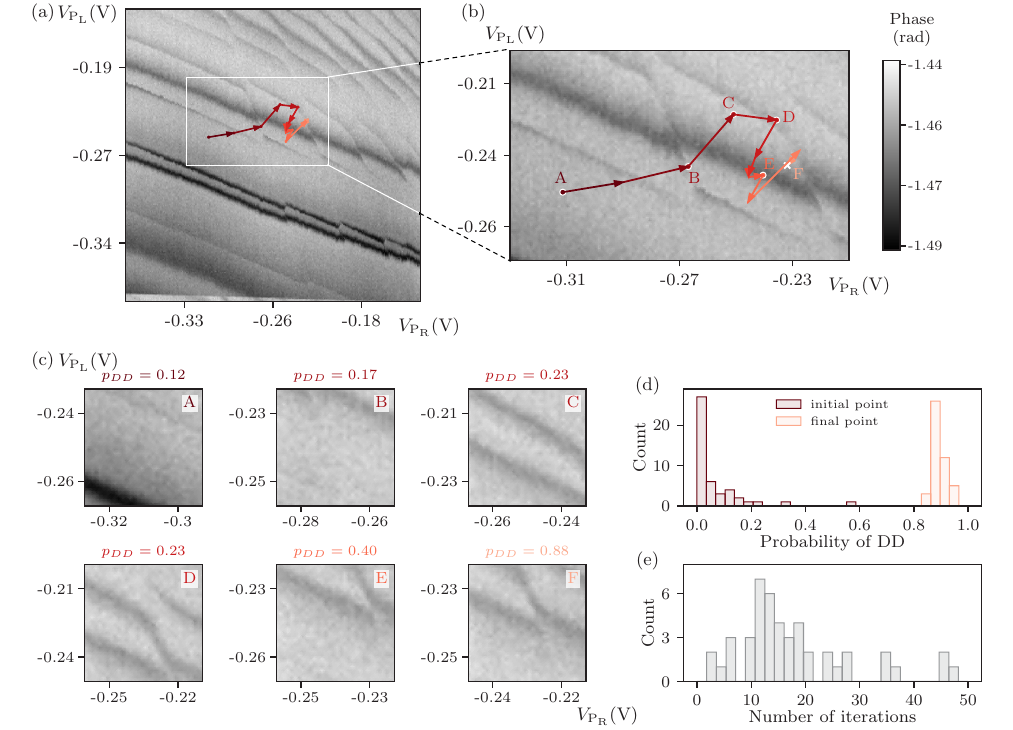}
    \centering
    \caption{
    (a) An overview of the sample run of the autotuning protocol implemented with active feedback in the larger space of plunger gates $(V_{\rm P_R}, V_{\rm P_L})$.
    (b) A close-up of a charge stability diagram showing a sample 12-step path of the coarse-tuning module with active feedback in the space of plunger gates $(V_{\rm P_R}, V_{\rm P_L})$.
    The arrows and the intensity of the color indicate the progress of the autotuner (dark to light).
    (c) The measured raw scans in the space of plunger gates $(V_{\rm P_R}, V_{\rm P_L})$ showing data available to the DSE module for six selected steps along the path shown in panel (a). 
    The p$_{DD}$ corresponds to the evaluated likelihood of the DD regime. 
    (d) The DD likelihood of initial and final points.
    (e) The distribution of path lengths across the 46 test runs.}
    \label{fig:coarse_tuning}
\end{figure*}

\begin{table}[b]
\renewcommand{\arraystretch}{1.05}
\renewcommand{\tabcolsep}{2pt}
\caption{\label{tab:boot_stats}
The summary statistics for pinch-off, saturation, and parking voltages for all finger gates assessed during the bootstrapping tests, averaged over $N_{\rm boot}=89$. 
}
\begin{ruledtabular}
\begin{tabular}{lrrr}
Gate & Pinch-off (\si{\milli\volt}) & Saturation (\si{\milli\volt}) & Parking (\si{\milli\volt}) \\ \hline
OB$_{\rm L}$ & $-762.9(2.1)$ & $-330.6(15.3)$ & $\cdots$ \\ 
P$_{\rm L}$ & $-539.0(1.3)$ & $-414.0(22.0)$ & $\cdots$ \\ 
B       & $-524.6(1.1)$ & $-500.4(12.4)$ & $\cdots$ \\ 
P$_{\rm R}$ & $-696.0(1.4)$ & $-463.7(17.6)$ & $\cdots$ \\ 
OB$_{\rm R}$ & $-525.2(0.9)$ & $-497.3(37.6)$ & $-507.5(1.4)$ \\ [5pt] 
SP & $-768.6(15.4)$ & $-344.6(8.1)$ & $-221.7(111)$  \\ [5pt]
BB & $\cdots$ & $\cdots$ & $-861.1(5.3)$ \\
SB$_{\rm T}$ & $\cdots$ & $\cdots$ & $-518.9(59.2)$ \\
SB$_{\rm B}$ & $\cdots$ & $\cdots$ & $-721.9(88.3)$ \\
\end{tabular}
\end{ruledtabular}
\end{table}

\section{Results} 
\label{sec:results}
The evaluation of the tuning algorithm is carried out in two phases. 
In the first phase, we test the performance of the individual steps of the bootstrapping algorithm. 
In the second phase, we perform a series of trial runs of the full autotuning algorithm in the $(V_{\rm P_R}, V_{\rm B}, V_{\rm P_L})$ space, with the B gate adjusted as described in Sec.~\ref{ssec:boot_QD_chan}.

\subsection{Bootstrapping: Performance analysis}
\label{ssec:boot_perf}
We initialize $N_{\rm boot}=89$ bootstrapping runs to test the performance of the autonomous bootstrapping algorithm presented in Fig. \ref{fig:algorithm-flow}(b). 
In each test run, we extract the results of 1D gate characterization as well as the final sensor-QD gate voltages obtained using rb measurements. 

The finger-gate characterization identifies working gates, and extracts pinch-off and saturation voltages. 
The five panels in the leftmost column in Fig.~\ref{fig:boot-stats}(a) identify on the device SEM the finger gates used to form and control qubit QDs: OB$_{\rm L}$, P$_{\rm L}$, B, P$_{\rm R}$, and OB$_{\rm R}$.
The corresponding 1D gate traces are shown in Fig.~\ref{fig:boot-stats}(b) and the extracted pinch-off and saturation voltages averaged over the $N_{\rm boot}=89$ tests are shown in Fig.~\ref{fig:boot-stats}(c). 
The error bars in Fig.~\ref{fig:boot-stats}(c) represent one standard deviation within the collected sample and provide a metric of algorithm and device stability.
High device and algorithm stability manifest through the low standard deviations of the extracted features: for the pinch-off, values range from $-762(2)$~\si{\milli\volt}~\footnote{We use a notation value(uncertainty) to express uncertainties, for example, $1.5(6)~\si{\centi\meter}$. 
All uncertainties herein reflect the uncorrelated combination of single-standard deviation statistical and systematic uncertainties.} for OB$_{\rm L}$ to $-525(1)$~\si{\milli\volt} for B; 
saturation values range from $-500(12)$~\si{\milli\volt} for B to $331(15)$~\si{\milli\volt} for OB$_{\rm L}$.
A summary of statistics for all finger gates is presented in Table~\ref{tab:boot_stats}. 

The middle column in Fig.~\ref{fig:boot-stats}(a) highlights the SP gate on the SEM.
An example of a 1D trace used to extract pinch-off and saturation voltages is shown in Fig.~\ref{fig:boot-stats}(b).
The average pinch-off and saturation values collected at the characterization stage, as well as the parking positions obtained by detecting the most sensitive position on the highest-scoring [according to the Eq.~(\ref{eq:2})] Coulomb peak in rf reflectometry after the formation of the sensor QD and the localization of the DD region is concluded, are depicted in Fig.~\ref{fig:boot-stats}(c). 
The standard deviations for pinch-off and saturation values of the SP gate extracted from current measurements are found to be $15$~\si{\milli\volt} and $8$~\si{\milli\volt}, respectively (for details, see Table~\ref{tab:boot_stats}). 
The standard deviation for the SP position selected on the highest-scoring Coulomb oscillation peak and parking at its steepest slope was at $111$~\si{\milli\volt}.
Since pinch-off values are more well-defined physically, with the slope being steeper than at saturation, we expected the standard deviation of saturation values to be higher than those of pinch-offs.

The two panels in the rightmost column in Fig.~\ref{fig:boot-stats}(a) show the pairs of gates used to form a sensor QD.
In Fig.~\ref{fig:boot-stats}(b), we show the analyzed rb measurements and in Fig.~\ref{fig:boot-stats}(c), we show the statistics for the parking voltages in (c). 
The parking voltages for the BB, SB$_{\rm T}$, and SB$_{\rm B}$ gates have been obtained using rb measurements, with the corresponding 1D traces shown in the rightmost panel in Fig.~\ref{fig:boot-stats}(b).
The parking voltage for the BB gate is the result of detecting a pinch-off in rb measurement taken in combination with OB$_{\rm R}$ [for performance statistics, see Fig.~\ref{fig:boot-stats}(c)]. 
The OB$_{\rm R}$ is the result of the succeeding fine tuning, as described in Sec.~\ref{ssec:boot_QD_chan}.
Finally, the parking positions of the sensor barriers correspond to the location of the most sensitive Coulomb oscillation peak [see Fig.~\ref{fig:boot-stats}(c)]. 
In this case, the dependence on the oscillatory behavior has resulted in a larger standard deviation of the parking voltages for the SB$_{\rm T}$ and SB$_{\rm B}$ gates, at 59~\si{\milli\volt} and 88~\si{\milli\volt}, respectively, than for BB and OB$_{\rm R}$, which rely on the well-behaving pinch-off values. 

To test the bootstrapping performance, we measure a large charge stability diagram in the $(V_{\rm P_R}, V_{\rm P_L})$ plunger space, with $V_{\rm B}=-500$~\si{\milli\volt} to qualitatively confirm that the qubit QDs can be sensed with the sensor QD.
The choice of $V_{\rm B}$ comes from prior work~\cite{Fedele19-PhD}.
Out of $89$ bootstrapping test runs, $12$ were unsuccessful at charge sensing the qubit QDs. 
Among these $12$ failed tuning attempts, $10$ had an uncharacteristically high parking voltage for the OB$_{\rm L}$ that provided an insufficient current pinch-off to form the left QD, and two remaining runs failed to detect the Coulomb oscillations in the rb measurements for SB$_{\rm T}$ versus SB$_{\rm B}$ measurements, and thus no sensor QD could be formed. 
The discovered failure modes have provided an experimental heuristic for invoking the sensor-QD re-tuning from the initial device state that was implemented for the full tuning protocol discussed in Sec.~\ref{ssec:full_perf}.

\subsection{Full tuning: performance analysis}
\label{ssec:full_perf}
Once the bootstrapping algorithm had been validated, the algorithm was expanded to include an active coarse-tuning feedback routine for calibrating the V$_{P_L}$, B, and V$_{P_R}$ gates to bring the QD device to a DD state [see Fig.~\ref{fig:algorithm-flow}(b)].  
The full tuning testing involves running the bootstrapping algorithm followed by the DSE-based coarse-tuning module as a single protocol, with the DSE module relying exclusively on rf reflectometry. 

The performance assessment for full tuning focuses on successful tuning to the DD state based on the outcome of the successful bootstrapping phase.
Given the random initialization of the coarse tuning, configurations of the individual gates after bootstrapping are re-used to perform multiple attempts at qubit DD tuning with randomized initial V$_{P_L}$, V$_{P_R}$, and B gate voltages. 
In cases when the DSE-based module reaches the initialization iteration limit,  the algorithm restarts bootstrapping.
A total of $N_{\rm full}=46$ complete runs are collected for further analysis.

In Fig.~\ref{fig:coarse_tuning}(a), we show an overview of a sample coarse-tuning run with active feedback in the $(V_{\rm P_R}, V_{\rm P_L})$ plunger gate space.
A close-up of the tuning path is shown in Fig.~\ref{fig:coarse_tuning}(b).
In this example, the QD device is initialized in a configuration corresponding to the SD$_{\rm L}$ state, with ${\rm\bf{p}}(V_\mathcal{R})=[0.2, 0.0, 0.0, 0.7, 0.1]$.
Then, through an iterative application of the DSE module along with the navigation routine, the configuration of the plunger gates is adjusted to bring the QD devices to the target DD state. 
In Fig.~\ref{fig:coarse_tuning}(c), we depict six $(38\times 38)-$\si{\milli \volt} scans with $(30\times30)$-pixels resolution taken along the tuning path, indicating an increased likelihood of the DD state as the algorithm navigates toward the DD state.

Out of $N_{\rm full}=46$ full tuning algorithm runs terminated at a gate configuration with the DD likelihood of the ${\rm\bf{p}}(V_\mathcal{R})$ state vector surpassing the predefined threshold of $85~\%$, two were qualitatively assessed as false positives. 
All remaining test runs were successful.
While human experts might have been able to recover a DD state from these failed positives, in the time spent recovering, the algorithm could possibly have tuned from scratch again. 
The probability distributions of the DD likelihood for the initial and the final state vectors are shown in Fig.~\ref{fig:coarse_tuning}(d).
The histogram in Fig.~\ref{fig:coarse_tuning}(e) shows the number of iterations it took for this algorithm to find the target state. 
We find a slight correlation between the probability of DD in the initial scan and the number of iterations.
In particular, while for test runs with $p_{\rm DD}^{\rm (init)}<0.11$ the number of iterations goes up to $N_{\rm iter}=47$, for $p_{\rm DD}^{\rm (init)}>0.11$, $N_{\rm iter}<21$.
The average time spent by the algorithm from starting bootstrapping until termination by successful tuning to the DD state was $7.7$~\si{\min}, with $5.7$~\si{\min} spent on bootstrapping and about $2~$\si{\min} on tuning. 
This compares very well with human expert tuning which, typically requires hours to reach charge sensing of double QDs~\cite{Moon20-ATQ}.

\section{Summary and outlook}
\label{sec:conclusion}
We propose and experimentally demonstrate a modular autonomous bootstrapping algorithm for bringing a GaAs spin-qubit QD device from a completely untuned state to the DD global-state configuration.
The algorithm performs gate characterization and device diagnostics, ensures the formation of a qubit-QD channel, and prepares the charge sensing QD for rf reflectometry.
The low initialization failure rate and high consistency of the saturation and pinch-off voltages confirm the stability of the algorithm.
The bootstrapping protocol efficiently deploys a combination of measurement techniques such as regular 1D voltage sweeps and rb measurements for the current sweeps and small and fast 2D sweeps using rf reflectometry, improving over previous demonstrations in the field that have used transport measurements. 
Moreover, the high success rate for tuning to the DD global-state configuration confirms the utility of the ML-based DSE module, trained exclusively using simulated data, for deployment across varying device architectures and materials.

The ability to perform autonomous tuning in charge sensing is especially important for dense arrays without access to individual transport measurements for each QD. 
The results presented in this work serve as a baseline for the development of a fully autonomous QD device initialization, calibration, and control system for large 1D and 2D QD arrays.
The next steps include incorporating data quality control modules to improve the reliability of the ML-based methods and integrating the charge-tuning and fine-tuning modules necessary for a full tune-up to the qubit stage. 
An important development will be the seamless integration of the QD device virtualization module to enable target control of qubit QDs.
Future work will also focus on extending first the bootstrapping and then the full algorithm to multiple QD arrays.

In the present work, the rf-reflectometry circuit (frequency, phase, position in the I-Q plane, demodulation parameters) has been set to known values optimized manually; in the future, the rf-reflectometry optimization could be automated using simple feedback based on measurements that are similar to those presented here.
Additionally, recent developments in high-speed and high-sensitivity real-time ``video mode'' data acquisition could be complemented by this protocol, allowing for more rapid sensor-QD tuning for charge sensing of the QD arrays, eliminating the need for human intervention. 
A broad application of our modules to different materials systems (e.g., nanowires, germanium quantum wells) might help us to understand differences between materials, disorders, and future platform-specific improvements.

Finally, this work focuses on GaAs devices, where a single layer of gates is sufficient to confine the QDs. 
This simplifies tuning compared to QDs defined by overlapping gate electrodes, which may add complexities due to cross-capacitance as well as a higher number of parameters that need to be adjusted.
While preliminary testing of some of the bootstrapping algorithm components shows good performance when tested on an accumulation-mode Si sample, full cross-platform compatibility will require the development of specialized data-analysis tools.

\begin{acknowledgments}
A.C. and T.R. acknowledge support from the Inge Lehmann Program of the Independent Research Fund Denmark. 
This work received funding from the U.S. Army Research Office (ARO) under Award No. W911NF-24-2-0043, the European Union Horizon 2020 Research and Innovation Programme under Grant Agreement No. 101017733 (QuantERA II), and through the Horizon Europe Framework Programme's Integrated Germanium Quantum Technology (IGNITE) project under Grant Agreement No. 101069515.
We thank Evert van Nieuwenberg for the valuable discussions and supervision of Lara Lausen.
The views and conclusions contained in this paper are those of the authors and should not be interpreted as representing the official policies, either expressed or implied, of the U.S. Government. 
The U.S. Government is authorized to reproduce and distribute reprints for Government purposes notwithstanding any copyright noted herein. 
Any mention of commercial products is for information only; it does not imply recommendation or endorsement by the National Institute of Standards and Technology (NIST).
\end{acknowledgments}

%


\begin{thebibliography}{32}%
\makeatletter
\providecommand \@ifxundefined [1]{%
 \@ifx{#1\undefined}
}%
\providecommand \@ifnum [1]{%
 \ifnum #1\expandafter \@firstoftwo
 \else \expandafter \@secondoftwo
 \fi
}%
\providecommand \@ifx [1]{%
 \ifx #1\expandafter \@firstoftwo
 \else \expandafter \@secondoftwo
 \fi
}%
\providecommand \natexlab [1]{#1}%
\providecommand \enquote  [1]{``#1''}%
\providecommand \bibnamefont  [1]{#1}%
\providecommand \bibfnamefont [1]{#1}%
\providecommand \citenamefont [1]{#1}%
\providecommand \href@noop [0]{\@secondoftwo}%
\providecommand \href [0]{\begingroup \@sanitize@url \@href}%
\providecommand \@href[1]{\@@startlink{#1}\@@href}%
\providecommand \@@href[1]{\endgroup#1\@@endlink}%
\providecommand \@sanitize@url [0]{\catcode `\\12\catcode `\$12\catcode
  `\&12\catcode `\#12\catcode `\^12\catcode `\_12\catcode `\%12\relax}%
\providecommand \@@startlink[1]{}%
\providecommand \@@endlink[0]{}%
\providecommand \url  [0]{\begingroup\@sanitize@url \@url }%
\providecommand \@url [1]{\endgroup\@href {#1}{\urlprefix }}%
\providecommand \urlprefix  [0]{URL }%
\providecommand \Eprint [0]{\href }%
\providecommand \doibase [0]{https://doi.org/}%
\providecommand \selectlanguage [0]{\@gobble}%
\providecommand \bibinfo  [0]{\@secondoftwo}%
\providecommand \bibfield  [0]{\@secondoftwo}%
\providecommand \translation [1]{[#1]}%
\providecommand \BibitemOpen [0]{}%
\providecommand \bibitemStop [0]{}%
\providecommand \bibitemNoStop [0]{.\EOS\space}%
\providecommand \EOS [0]{\spacefactor3000\relax}%
\providecommand \BibitemShut  [1]{\csname bibitem#1\endcsname}%
\let\auto@bib@innerbib\@empty
\bibitem {Loss98-QCD}%
  \BibitemOpen
  \bibfield  {author} {\bibinfo {author} {\bibfnamefont {D.}~\bibnamefont
  {Loss}}\ and\ \bibinfo {author} {\bibfnamefont {D.~P.}\ \bibnamefont
  {DiVincenzo}},\ }\bibfield  {title} {\bibinfo {title} {Quantum computation
  with quantum dots},\ }\href {https://doi.org/10.1103/PhysRevA.57.120}
  {\bibfield  {journal} {\bibinfo  {journal} {Phys. Rev. A}\ }\textbf {\bibinfo
  {volume} {57}},\ \bibinfo {pages} {120} (\bibinfo {year} {1998})}\BibitemShut
  {NoStop}%
%
\bibitem {Vandersypen19-QCS}%
  \BibitemOpen
  \bibfield  {author} {\bibinfo {author} {\bibfnamefont {L.~M.~K.}\
  \bibnamefont {Vandersypen}}\ and\ \bibinfo {author} {\bibfnamefont {M.~A.}\
  \bibnamefont {Eriksson}},\ }\bibfield  {title} {\bibinfo {title} {Quantum
  computing with semiconductor spins},\ }\href
  {https://doi.org/doi:10.1063/PT.3.4270} {\bibfield  {journal} {\bibinfo
  {journal} {Phys. Today}\ }\textbf {\bibinfo {volume} {72}},\ \bibinfo {pages}
  {38} (\bibinfo {year} {2019})}\BibitemShut {NoStop}%
%
\bibitem {Bavdaz22-QDC}%
  \BibitemOpen
  \bibfield  {author} {\bibinfo {author} {\bibfnamefont {P.~L.}\ \bibnamefont
  {Bavdaz}}, \bibinfo {author} {\bibfnamefont {H.~G.~J.}\ \bibnamefont
  {Eenink}}, \bibinfo {author} {\bibfnamefont {J.}~\bibnamefont {van
  Staveren}}, \bibinfo {author} {\bibfnamefont {M.}~\bibnamefont {Lodari}},
  \bibinfo {author} {\bibfnamefont {C.~G.}\ \bibnamefont {Almudever}}, \bibinfo
  {author} {\bibfnamefont {J.~S.}\ \bibnamefont {Clarke}}, \bibinfo {author}
  {\bibfnamefont {F.}~\bibnamefont {Sebasatiano}}, \bibinfo {author}
  {\bibfnamefont {M.}~\bibnamefont {Veldhorst}},\ and\ \bibinfo {author}
  {\bibfnamefont {G.}~\bibnamefont {Scappucci}},\ }\bibfield  {title} {\bibinfo
  {title} {A quantum dot crossbar array with sublinear scaling of interconnects
  at cryogenic temperature},\ }\href
  {https://doi.org/10.1038/s41534-022-00597-1} {\bibfield  {journal} {\bibinfo
  {journal} {npj Quantum Inf.}\ }\textbf {\bibinfo {volume} {8}},\ \bibinfo
  {pages} {1} (\bibinfo {year} {2022})}\BibitemShut {NoStop}%
%
\bibitem {Boter22-SWA}%
  \BibitemOpen
  \bibfield  {author} {\bibinfo {author} {\bibfnamefont {J.~M.}\ \bibnamefont
  {Boter}}, \bibinfo {author} {\bibfnamefont {J.~P.}\ \bibnamefont
  {Dehollain}}, \bibinfo {author} {\bibfnamefont {J.~P.}\ \bibnamefont {van
  Dijk}}, \bibinfo {author} {\bibfnamefont {Y.}~\bibnamefont {Xu}}, \bibinfo
  {author} {\bibfnamefont {T.}~\bibnamefont {Hensgens}}, \bibinfo {author}
  {\bibfnamefont {R.}~\bibnamefont {Versluis}}, \bibinfo {author}
  {\bibfnamefont {H.~W.}\ \bibnamefont {Naus}}, \bibinfo {author}
  {\bibfnamefont {J.~S.}\ \bibnamefont {Clarke}}, \bibinfo {author}
  {\bibfnamefont {M.}~\bibnamefont {Veldhorst}}, \bibinfo {author}
  {\bibfnamefont {F.}~\bibnamefont {Sebastiano}},\ and\ \bibinfo {author}
  {\bibfnamefont {L.~M.}\ \bibnamefont {Vandersypen}},\ }\bibfield  {title}
  {\bibinfo {title} {Spiderweb array: A sparse spin-qubit array},\ }\href
  {https://doi.org/10.1103/PhysRevApplied.18.024053} {\bibfield  {journal}
  {\bibinfo  {journal} {Phys. Rev. Applied}\ }\textbf {\bibinfo {volume}
  {18}},\ \bibinfo {pages} {024053} (\bibinfo {year} {2022})}\BibitemShut
  {NoStop}%
%
\bibitem {Ansaloni20-FFQ}%
  \BibitemOpen
  \bibfield  {author} {\bibinfo {author} {\bibfnamefont {F.}~\bibnamefont
  {Ansaloni}}, \bibinfo {author} {\bibfnamefont {A.}~\bibnamefont
  {Chatterjee}}, \bibinfo {author} {\bibfnamefont {H.}~\bibnamefont
  {Bohuslavskyi}}, \bibinfo {author} {\bibfnamefont {B.}~\bibnamefont
  {Bertrand}}, \bibinfo {author} {\bibfnamefont {L.}~\bibnamefont {Hutin}},
  \bibinfo {author} {\bibfnamefont {M.}~\bibnamefont {Vinet}},\ and\ \bibinfo
  {author} {\bibfnamefont {F.}~\bibnamefont {Kuemmeth}},\ }\bibfield  {title}
  {\bibinfo {title} {Single-electron operations in a foundry-fabricated array
  of quantum dots},\ }\href {https://doi.org/10.1038/s41467-020-20280-3}
  {\bibfield  {journal} {\bibinfo  {journal} {Nat. Commun.}\ }\textbf {\bibinfo
  {volume} {11}},\ \bibinfo {pages} {6399} (\bibinfo {year}
  {2020})}\BibitemShut {NoStop}%
%
\bibitem {Watson18-TQP}%
  \BibitemOpen
  \bibfield  {author} {\bibinfo {author} {\bibfnamefont {T.~F.}\ \bibnamefont
  {Watson}}, \bibinfo {author} {\bibfnamefont {S.~G.~J.}\ \bibnamefont
  {Philips}}, \bibinfo {author} {\bibfnamefont {E.}~\bibnamefont {Kawakami}},
  \bibinfo {author} {\bibfnamefont {D.~R.}\ \bibnamefont {Ward}}, \bibinfo
  {author} {\bibfnamefont {P.}~\bibnamefont {Scarlino}}, \bibinfo {author}
  {\bibfnamefont {M.}~\bibnamefont {Veldhorst}}, \bibinfo {author}
  {\bibfnamefont {D.~E.}\ \bibnamefont {Savage}}, \bibinfo {author}
  {\bibfnamefont {M.~G.}\ \bibnamefont {Lagally}}, \bibinfo {author}
  {\bibfnamefont {M.}~\bibnamefont {Friesen}}, \bibinfo {author} {\bibfnamefont
  {S.~N.}\ \bibnamefont {Coppersmith}}, \bibinfo {author} {\bibfnamefont
  {M.~A.}\ \bibnamefont {Eriksson}},\ and\ \bibinfo {author} {\bibfnamefont
  {L.~M.~K.}\ \bibnamefont {Vandersypen}},\ }\bibfield  {title} {\bibinfo
  {title} {A programmable two-qubit quantum processor in silicon},\ }\href
  {https://doi.org/10.1038/nature25766} {\bibfield  {journal} {\bibinfo
  {journal} {Nature}\ }\textbf {\bibinfo {volume} {555}},\ \bibinfo {pages}
  {633} (\bibinfo {year} {2018})}\BibitemShut {NoStop}%
%
\bibitem {Xue21-CCC}%
  \BibitemOpen
  \bibfield  {author} {\bibinfo {author} {\bibfnamefont {X.}~\bibnamefont
  {Xue}}, \bibinfo {author} {\bibfnamefont {B.}~\bibnamefont {Patra}}, \bibinfo
  {author} {\bibfnamefont {J.~P.~G.}\ \bibnamefont {van Dijk}}, \bibinfo
  {author} {\bibfnamefont {N.}~\bibnamefont {Samkharadze}}, \bibinfo {author}
  {\bibfnamefont {S.}~\bibnamefont {Subramanian}}, \bibinfo {author}
  {\bibfnamefont {A.}~\bibnamefont {Corna}}, \bibinfo {author} {\bibfnamefont
  {B.}~\bibnamefont {Paquelet~Wuetz}}, \bibinfo {author} {\bibfnamefont
  {C.}~\bibnamefont {Jeon}}, \bibinfo {author} {\bibfnamefont {F.}~\bibnamefont
  {Sheikh}}, \bibinfo {author} {\bibfnamefont {E.}~\bibnamefont
  {Juarez-Hernandez}}, \emph {et~al.}\ }\bibfield  {title} {\bibinfo
  {title} {Cmos-based cryogenic control of silicon quantum circuits},\ }\href
  {https://doi.org/10.1038/s41586-021-03469-4} {\bibfield  {journal} {\bibinfo
  {journal} {Nature}\ }\textbf {\bibinfo {volume} {593}},\ \bibinfo {pages}
  {205} (\bibinfo {year} {2021})}\BibitemShut {NoStop}%
%
\bibitem {Zwerver22-QMA}%
  \BibitemOpen
  \bibfield  {author} {\bibinfo {author} {\bibfnamefont {A.~M.~J.}\
  \bibnamefont {Zwerver}}, \bibinfo {author} {\bibfnamefont {T.}~\bibnamefont
  {Krähenmann}}, \bibinfo {author} {\bibfnamefont {T.~F.}\ \bibnamefont
  {Watson}}, \bibinfo {author} {\bibfnamefont {L.}~\bibnamefont {Lampert}},
  \bibinfo {author} {\bibfnamefont {H.~C.}\ \bibnamefont {George}}, \bibinfo
  {author} {\bibfnamefont {R.}~\bibnamefont {Pillarisetty}}, \bibinfo {author}
  {\bibfnamefont {S.~A.}\ \bibnamefont {Bojarski}}, \bibinfo {author}
  {\bibfnamefont {P.}~\bibnamefont {Amin}}, \bibinfo {author} {\bibfnamefont
  {S.~V.}\ \bibnamefont {Amitonov}}, \bibinfo {author} {\bibfnamefont {J.~M.}\
  \bibnamefont {Boter}}, \emph{et al.} \ }\bibfield  {title} {\bibinfo {title} {Qubits made
  by advanced semiconductor manufacturing},\ }\href
  {https://doi.org/10.1038/s41928-022-00727-9} {\bibfield  {journal} {\bibinfo
  {journal} {Nat. Electron.}\ }\textbf {\bibinfo {volume} {5}},\ \bibinfo
  {pages} {184} (\bibinfo {year} {2022})}\BibitemShut {NoStop}%
%
\bibitem {Noiri22-FUG}%
  \BibitemOpen
  \bibfield  {author} {\bibinfo {author} {\bibfnamefont {A.}~\bibnamefont
  {Noiri}}, \bibinfo {author} {\bibfnamefont {K.}~\bibnamefont {Takeda}},
  \bibinfo {author} {\bibfnamefont {T.}~\bibnamefont {Nakajima}}, \bibinfo
  {author} {\bibfnamefont {T.}~\bibnamefont {Kobayashi}}, \bibinfo {author}
  {\bibfnamefont {A.}~\bibnamefont {Sammak}}, \bibinfo {author} {\bibfnamefont
  {G.}~\bibnamefont {Scappucci}},\ and\ \bibinfo {author} {\bibfnamefont
  {S.}~\bibnamefont {Tarucha}},\ }\bibfield  {title} {\bibinfo {title} {Fast
  universal quantum gate above the fault-tolerance threshold in silicon},\
  }\href {https://doi.org/10.1038/s41586-021-04182-y} {\bibfield  {journal}
  {\bibinfo  {journal} {Nature}\ }\textbf {\bibinfo {volume} {601}},\ \bibinfo
  {pages} {338} (\bibinfo {year} {2022})}\BibitemShut {NoStop}%
%
\bibitem {Philips22-UCS}%
  \BibitemOpen
  \bibfield  {author} {\bibinfo {author} {\bibfnamefont {S.~G.~J.}\
  \bibnamefont {Philips}}, \bibinfo {author} {\bibfnamefont {M.~T.}\
  \bibnamefont {Madzik}}, \bibinfo {author} {\bibfnamefont {S.~V.}\
  \bibnamefont {Amitonov}}, \bibinfo {author} {\bibfnamefont {S.~L.}\
  \bibnamefont {de~Snoo}}, \bibinfo {author} {\bibfnamefont {M.}~\bibnamefont
  {Russ}}, \bibinfo {author} {\bibfnamefont {N.}~\bibnamefont {Kalhor}},
  \bibinfo {author} {\bibfnamefont {C.}~\bibnamefont {Volk}}, \bibinfo {author}
  {\bibfnamefont {W.~I.~L.}\ \bibnamefont {Lawrie}}, \bibinfo {author}
  {\bibfnamefont {D.}~\bibnamefont {Brousse}}, \bibinfo {author} {\bibfnamefont
  {L.}~\bibnamefont {Tryputen}}, \bibinfo {author} {\bibfnamefont {B.~P.}\
  \bibnamefont {Wuetz}}, \bibinfo {author} {\bibfnamefont {A.}~\bibnamefont
  {Sammak}}, \bibinfo {author} {\bibfnamefont {M.}~\bibnamefont {Veldhorst}},
  \bibinfo {author} {\bibfnamefont {G.}~\bibnamefont {Scappucci}},\ and\
  \bibinfo {author} {\bibfnamefont {L.~M.~K.}\ \bibnamefont {Vandersypen}},\
  }\bibfield  {title} {\bibinfo {title} {Universal control of a six-qubit
  quantum processor in silicon},\ }\href
  {https://doi.org/10.1038/s41586-022-05117-x} {\bibfield  {journal} {\bibinfo
  {journal} {Nature}\ }\textbf {\bibinfo {volume} {609}},\ \bibinfo {pages}
  {919} (\bibinfo {year} {2022})}\BibitemShut {NoStop}%
%
\bibitem {Weinstein22-ULS}%
  \BibitemOpen
  \bibfield  {author} {\bibinfo {author} {\bibfnamefont {A.~J.}\ \bibnamefont
  {Weinstein}}, \bibinfo {author} {\bibfnamefont {M.~D.}\ \bibnamefont {Reed}},
  \bibinfo {author} {\bibfnamefont {A.~M.}\ \bibnamefont {Jones}}, \bibinfo
  {author} {\bibfnamefont {R.~W.}\ \bibnamefont {Andrews}}, \bibinfo {author}
  {\bibfnamefont {D.}~\bibnamefont {Barnes}}, \bibinfo {author} {\bibfnamefont
  {J.~Z.}\ \bibnamefont {Blumoff}}, \bibinfo {author} {\bibfnamefont {L.~E.}\
  \bibnamefont {Euliss}}, \bibinfo {author} {\bibfnamefont {K.}~\bibnamefont
  {Eng}}, \bibinfo {author} {\bibfnamefont {B.}~\bibnamefont {Fong}}, \bibinfo
  {author} {\bibfnamefont {S.~D.}\ \bibnamefont {Ha}}, \emph{et al.} \ }\bibfield  {title} {\bibinfo {title} {Universal
  logic with encoded spin qubits in silicon},\ }\href
  {https://doi.org/10.1038/s41586-023-05777-3} {\bibfield  {journal} {\bibinfo
  {journal} {Nature}\ }\textbf {\bibinfo {volume} {615}},\ \bibinfo {pages}
  {817} (\bibinfo {year} {2023})}\BibitemShut {NoStop}%
%
\bibitem {Takeda22-QES}%
  \BibitemOpen
  \bibfield  {author} {\bibinfo {author} {\bibfnamefont {K.}~\bibnamefont
  {Takeda}}, \bibinfo {author} {\bibfnamefont {A.}~\bibnamefont {Noiri}},
  \bibinfo {author} {\bibfnamefont {T.}~\bibnamefont {Nakajima}}, \bibinfo
  {author} {\bibfnamefont {T.}~\bibnamefont {Kobayashi}},\ and\ \bibinfo
  {author} {\bibfnamefont {S.}~\bibnamefont {Tarucha}},\ }\bibfield  {title}
  {\bibinfo {title} {Quantum error correction with silicon spin qubits},\
  }\href {https://doi.org/10.1038/s41586-022-04986-6} {\bibfield  {journal}
  {\bibinfo  {journal} {Nature}\ }\textbf {\bibinfo {volume} {608}},\ \bibinfo
  {pages} {682} (\bibinfo {year} {2022})}\BibitemShut {NoStop}%
%
\bibitem {Madzik22-PTT}%
  \BibitemOpen
  \bibfield  {author} {\bibinfo {author} {\bibfnamefont {M.~T.}\ \bibnamefont
  {Madzik}}, \bibinfo {author} {\bibfnamefont {S.}~\bibnamefont {Asaad}},
  \bibinfo {author} {\bibfnamefont {A.}~\bibnamefont {Youssry}}, \bibinfo
  {author} {\bibfnamefont {B.}~\bibnamefont {Joecker}}, \bibinfo {author}
  {\bibfnamefont {K.~M.}\ \bibnamefont {Rudinger}}, \bibinfo {author}
  {\bibfnamefont {E.}~\bibnamefont {Nielsen}}, \bibinfo {author} {\bibfnamefont
  {K.~C.}\ \bibnamefont {Young}}, \bibinfo {author} {\bibfnamefont {T.~J.}\
  \bibnamefont {Proctor}}, \bibinfo {author} {\bibfnamefont {A.~D.}\
  \bibnamefont {Baczewski}}, \bibinfo {author} {\bibfnamefont {A.}~\bibnamefont
  {Laucht}}, \emph{et al.} \ }\bibfield  {title} {\bibinfo {title}
  {Precision tomography of a three-qubit donor quantum processor in silicon},\
  }\href {https://doi.org/10.1038/s41586-021-04292-7} {\bibfield  {journal}
  {\bibinfo  {journal} {Nature}\ }\textbf {\bibinfo {volume} {601}},\ \bibinfo
  {pages} {348} (\bibinfo {year} {2022})}\BibitemShut {NoStop}%
%
\bibitem {Chatterjee21-AEC}%
  \BibitemOpen
  \bibfield  {author} {\bibinfo {author} {\bibfnamefont {A.}~\bibnamefont
  {Chatterjee}}, \bibinfo {author} {\bibfnamefont {F.}~\bibnamefont
  {Ansaloni}}, \bibinfo {author} {\bibfnamefont {T.}~\bibnamefont {Rasmussen}},
  \bibinfo {author} {\bibfnamefont {B.}~\bibnamefont {Brovang}}, \bibinfo
  {author} {\bibfnamefont {F.}~\bibnamefont {Fedele}}, \bibinfo {author}
  {\bibfnamefont {H.}~\bibnamefont {Bohuslavskyi}}, \bibinfo {author}
  {\bibfnamefont {O.}~\bibnamefont {Krause}},\ and\ \bibinfo {author}
  {\bibfnamefont {F.}~\bibnamefont {Kuemmeth}},\ }\bibfield  {title} {\bibinfo
  {title} {Autonomous estimation of high-dimensional coulomb diamonds from
  sparse measurements},\ }\href
  {https://doi.org/10.1103/PhysRevApplied.18.064040} {\bibfield  {journal}
  {\bibinfo  {journal} {Phys. Rev. Appl.}\ }\textbf {\bibinfo {volume} {18}},\
  \bibinfo {pages} {064040} (\bibinfo {year} {2022})}\BibitemShut {NoStop}%
%
\bibitem {Burkard21-SSQ}%
  \BibitemOpen
  \bibfield  {author} {\bibinfo {author} {\bibfnamefont {G.}~\bibnamefont
  {Burkard}}, \bibinfo {author} {\bibfnamefont {T.~D.}\ \bibnamefont {Ladd}},
  \bibinfo {author} {\bibfnamefont {A.}~\bibnamefont {Pan}}, \bibinfo {author}
  {\bibfnamefont {J.~M.}\ \bibnamefont {Nichol}},\ and\ \bibinfo {author}
  {\bibfnamefont {J.~R.}\ \bibnamefont {Petta}},\ }\bibfield  {title} {\bibinfo
  {title} {Semiconductor spin qubits},\ }\href
  {https://doi.org/10.1103/RevModPhys.95.025003} {\bibfield  {journal}
  {\bibinfo  {journal} {Rev. Mod. Phys.}\ }\textbf {\bibinfo {volume} {95}},\
  \bibinfo {pages} {025003} (\bibinfo {year} {2023})}\BibitemShut {NoStop}%
%
\bibitem {Zwolak21-AAQ}%
  \BibitemOpen
  \bibfield  {author} {\bibinfo {author} {\bibfnamefont {J.~P.}\ \bibnamefont
  {Zwolak}}\ and\ \bibinfo {author} {\bibfnamefont {J.~M.}\ \bibnamefont
  {Taylor}},\ }\bibfield  {title} {\bibinfo {title} {{\it Colloquium}: Advances
  in automation of quantum dot devices control},\ }\href
  {https://doi.org/10.1103/RevModPhys.95.011006} {\bibfield  {journal}
  {\bibinfo  {journal} {Rev. Mod. Phys.}\ }\textbf {\bibinfo {volume} {95}},\
  \bibinfo {pages} {011006} (\bibinfo {year} {2023})}\BibitemShut {NoStop}%
%
\bibitem {Ziegler22-TRA}%
  \BibitemOpen
  \bibfield  {author} {\bibinfo {author} {\bibfnamefont {J.}~\bibnamefont
  {Ziegler}}, \bibinfo {author} {\bibfnamefont {T.}~\bibnamefont {McJunkin}},
  \bibinfo {author} {\bibfnamefont {E.~S.}\ \bibnamefont {Joseph}}, \bibinfo
  {author} {\bibfnamefont {S.~S.}\ \bibnamefont {Kalantre}}, \bibinfo {author}
  {\bibfnamefont {B.}~\bibnamefont {Harpt}}, \bibinfo {author} {\bibfnamefont
  {D.~E.}\ \bibnamefont {Savage}}, \bibinfo {author} {\bibfnamefont {M.~G.}\
  \bibnamefont {Lagally}}, \bibinfo {author} {\bibfnamefont {M.~A.}\
  \bibnamefont {Eriksson}}, \bibinfo {author} {\bibfnamefont {J.~M.}\
  \bibnamefont {Taylor}},\ and\ \bibinfo {author} {\bibfnamefont {J.~P.}\
  \bibnamefont {Zwolak}},\ }\bibfield  {title} {\bibinfo {title} {Toward robust
  autotuning of noisy quantum dot devices},\ }\href
  {https://doi.org/10.1103/PhysRevApplied.17.024069} {\bibfield  {journal}
  {\bibinfo  {journal} {Phys. Rev. Appl.}\ }\textbf {\bibinfo {volume} {17}},\
  \bibinfo {pages} {024069} (\bibinfo {year} {2022})}\BibitemShut {NoStop}%
%
\bibitem {Zwolak20-AQD}%
  \BibitemOpen
  \bibfield  {author} {\bibinfo {author} {\bibfnamefont {J.~P.}\ \bibnamefont
  {Zwolak}}, \bibinfo {author} {\bibfnamefont {T.}~\bibnamefont {McJunkin}},
  \bibinfo {author} {\bibfnamefont {S.~S.}\ \bibnamefont {Kalantre}}, \bibinfo
  {author} {\bibfnamefont {J.}~\bibnamefont {Dodson}}, \bibinfo {author}
  {\bibfnamefont {E.~R.}\ \bibnamefont {MacQuarrie}}, \bibinfo {author}
  {\bibfnamefont {D.}~\bibnamefont {Savage}}, \bibinfo {author} {\bibfnamefont
  {M.}~\bibnamefont {Lagally}}, \bibinfo {author} {\bibfnamefont
  {S.}~\bibnamefont {Coppersmith}}, \bibinfo {author} {\bibfnamefont {M.~A.}\
  \bibnamefont {Eriksson}},\ and\ \bibinfo {author} {\bibfnamefont {J.~M.}\
  \bibnamefont {Taylor}},\ }\bibfield  {title} {\bibinfo {title} {Autotuning of
  double-dot devices in situ with machine learning},\ }\href
  {https://doi.org/10.1103/PhysRevApplied.13.034075} {\bibfield  {journal}
  {\bibinfo  {journal} {Phys. Rev. Appl.}\ }\textbf {\bibinfo {volume} {13}},\
  \bibinfo {pages} {034075} (\bibinfo {year} {2020})}\BibitemShut {NoStop}%
%
\bibitem {Ziegler22-TAR}%
  \BibitemOpen
  \bibfield  {author} {\bibinfo {author} {\bibfnamefont {J.}~\bibnamefont
  {Ziegler}}, \bibinfo {author} {\bibfnamefont {F.}~\bibnamefont {Luthi}},
  \bibinfo {author} {\bibfnamefont {M.}~\bibnamefont {Ramsey}}, \bibinfo
  {author} {\bibfnamefont {F.}~\bibnamefont {Borjans}}, \bibinfo {author}
  {\bibfnamefont {G.}~\bibnamefont {Zheng}},\ and\ \bibinfo {author}
  {\bibfnamefont {J.~P.}\ \bibnamefont {Zwolak}},\ }\bibfield  {title}
  {\bibinfo {title} {Tuning arrays with rays: Physics-informed tuning of
  quantum dot charge states},\ }\href
  {https://doi.org/10.1103/PhysRevApplied.20.034067} {\bibfield  {journal}
  {\bibinfo  {journal} {Phys. Rev. Appl.}\ }\textbf {\bibinfo {volume} {20}},\
  \bibinfo {pages} {034067} (\bibinfo {year} {2023}{\natexlab{a}})}\BibitemShut
  {NoStop}%
%
\bibitem {Zwolak21-RBI}%
  \BibitemOpen
  \bibfield  {author} {\bibinfo {author} {\bibfnamefont {J.~P.}\ \bibnamefont
  {Zwolak}}, \bibinfo {author} {\bibfnamefont {T.}~\bibnamefont {McJunkin}},
  \bibinfo {author} {\bibfnamefont {S.~S.}\ \bibnamefont {Kalantre}}, \bibinfo
  {author} {\bibfnamefont {S.~F.}\ \bibnamefont {Neyens}}, \bibinfo {author}
  {\bibfnamefont {E.~R.}\ \bibnamefont {MacQuarrie}}, \bibinfo {author}
  {\bibfnamefont {M.~A.}\ \bibnamefont {Eriksson}},\ and\ \bibinfo {author}
  {\bibfnamefont {J.~M.}\ \bibnamefont {Taylor}},\ }\bibfield  {title}
  {\bibinfo {title} {Ray-based framework for state identification in quantum
  dot devices},\ }\href {https://doi.org/10.1103/PRXQuantum.2.020335}
  {\bibfield  {journal} {\bibinfo  {journal} {PRX Quantum}\ }\textbf {\bibinfo
  {volume} {2}},\ \bibinfo {pages} {020335} (\bibinfo {year}
  {2021})}\BibitemShut {NoStop}%
%
\bibitem {Berritta23-RTC}%
  \BibitemOpen
  \bibfield  {author} {\bibinfo {author} {\bibfnamefont {F.}~\bibnamefont
  {Berritta}}, \bibinfo {author} {\bibfnamefont {T.}~\bibnamefont {Rasmussen}},
  \bibinfo {author} {\bibfnamefont {J.~A.}\ \bibnamefont {Krzywda}}, \bibinfo
  {author} {\bibfnamefont {J.}~\bibnamefont {van~der Heijden}}, \bibinfo
  {author} {\bibfnamefont {F.}~\bibnamefont {Fedele}}, \bibinfo {author}
  {\bibfnamefont {S.}~\bibnamefont {Fallahi}}, \bibinfo {author} {\bibfnamefont
  {G.~C.}\ \bibnamefont {Gardner}}, \bibinfo {author} {\bibfnamefont {M.~J.}\
  \bibnamefont {Manfra}}, \bibinfo {author} {\bibfnamefont {E.}~\bibnamefont
  {van Nieuwenburg}}, \bibinfo {author} {\bibfnamefont {J.}~\bibnamefont
  {Danon}}, \bibinfo {author} {\bibfnamefont {A.}~\bibnamefont {Chatterjee}},\
  and\ \bibinfo {author} {\bibfnamefont {F.}~\bibnamefont {Kuemmeth}},\
  }\bibfield  {title} {\bibinfo {title} {Real-time two-axis control of a spin
  qubit},\ }\href {https://doi.org/10.1038/s41467-024-45857-0} {\bibfield
  {journal} {\bibinfo  {journal} {Nat. Commun.}\ }\textbf {\bibinfo {volume}
  {15}},\ \bibinfo {pages} {1676} (\bibinfo {year} {2024})}\BibitemShut
  {NoStop}%
%
\bibitem {Hickie24-ALC}%
  \BibitemOpen
  \bibfield  {author} {\bibinfo {author} {\bibfnamefont {J.}~\bibnamefont
  {Hickie}}, \bibinfo {author} {\bibfnamefont {B.}~\bibnamefont {van
  Straaten}}, \bibinfo {author} {\bibfnamefont {F.}~\bibnamefont {Fedele}},
  \bibinfo {author} {\bibfnamefont {D.}~\bibnamefont {Jirovec}}, \bibinfo
  {author} {\bibfnamefont {A.}~\bibnamefont {Ballabio}}, \bibinfo {author}
  {\bibfnamefont {D.}~\bibnamefont {Chrastina}}, \bibinfo {author}
  {\bibfnamefont {G.}~\bibnamefont {Isella}}, \bibinfo {author} {\bibfnamefont
  {G.}~\bibnamefont {Katsaros}},\ and\ \bibinfo {author} {\bibfnamefont
  {N.}~\bibnamefont {Ares}},\ }\bibfield  {title} {\bibinfo {title} {Automated
  long-range compensation of an rf quantum dot sensor},\ }\href
  {https://doi.org/10.1103/PhysRevApplied.22.064026} {\bibfield  {journal}
  {\bibinfo  {journal} {Phys. Rev. Appl.}\ }\textbf {\bibinfo {volume} {22}},\
  \bibinfo {pages} {064026} (\bibinfo {year} {2024})}\BibitemShut {NoStop}%
%
\bibitem {Moon20-ATQ}%
  \BibitemOpen
  \bibfield  {author} {\bibinfo {author} {\bibfnamefont {H.}~\bibnamefont
  {Moon}}, \bibinfo {author} {\bibfnamefont {D.~T.}\ \bibnamefont {Lennon}},
  \bibinfo {author} {\bibfnamefont {J.}~\bibnamefont {Kirkpatrick}}, \bibinfo
  {author} {\bibfnamefont {N.~M.}\ \bibnamefont {van Esbroeck}}, \bibinfo
  {author} {\bibfnamefont {L.~C.}\ \bibnamefont {Camenzind}}, \bibinfo {author}
  {\bibfnamefont {L.}~\bibnamefont {Yu}}, \bibinfo {author} {\bibfnamefont
  {F.}~\bibnamefont {Vigneau}}, \bibinfo {author} {\bibfnamefont {D.~M.}\
  \bibnamefont {Zumb\"uhl}}, \bibinfo {author} {\bibfnamefont {G.~A.~D.}\
  \bibnamefont {Briggs}}, \bibinfo {author} {\bibfnamefont {M.~A.}\
  \bibnamefont {Osborne}}, \bibinfo {author} {\bibfnamefont {D.}~\bibnamefont
  {Sejdinovic}}, \bibinfo {author} {\bibfnamefont {E.~A.}\ \bibnamefont
  {Laird}},\ and\ \bibinfo {author} {\bibfnamefont {N.}~\bibnamefont {Ares}},\
  }\bibfield  {title} {\bibinfo {title} {Machine learning enables completely
  automatic tuning of a quantum device faster than human experts},\ }\href
  {https://doi.org/10.1038/s41467-020-17835-9} {\bibfield  {journal} {\bibinfo
  {journal} {Nat. Commun.}\ }\textbf {\bibinfo {volume} {11}},\ \bibinfo
  {pages} {4161} (\bibinfo {year} {2020})}\BibitemShut {NoStop}%
%
\bibitem {Podd10-CSQ}%
  \BibitemOpen
  \bibfield  {author} {\bibinfo {author} {\bibfnamefont {G.~J.}\ \bibnamefont
  {Podd}}, \bibinfo {author} {\bibfnamefont {S.~J.}\ \bibnamefont {Angus}},
  \bibinfo {author} {\bibfnamefont {D.~A.}\ \bibnamefont {Williams}},\ and\
  \bibinfo {author} {\bibfnamefont {A.~J.}\ \bibnamefont {Ferguson}},\
  }\bibfield  {title} {\bibinfo {title} {Charge sensing in intrinsic silicon
  quantum dot},\ }\href {https://doi.org/10.1063/1.3318463} {\bibfield
  {journal} {\bibinfo  {journal} {Appl. Phys. Lett.}\ }\textbf {\bibinfo
  {volume} {96}},\ \bibinfo {pages} {082104} (\bibinfo {year}
  {2010})}\BibitemShut {NoStop}%
%
\bibitem {Durrer19-ATQ}%
  \BibitemOpen
  \bibfield  {author} {\bibinfo {author} {\bibfnamefont {R.}~\bibnamefont
  {Durrer}}, \bibinfo {author} {\bibfnamefont {B.}~\bibnamefont {Kratochwil}},
  \bibinfo {author} {\bibfnamefont {J.}~\bibnamefont {Koski}}, \bibinfo
  {author} {\bibfnamefont {A.}~\bibnamefont {Landig}}, \bibinfo {author}
  {\bibfnamefont {C.}~\bibnamefont {Reichl}}, \bibinfo {author} {\bibfnamefont
  {W.}~\bibnamefont {Wegscheider}}, \bibinfo {author} {\bibfnamefont
  {T.}~\bibnamefont {Ihn}},\ and\ \bibinfo {author} {\bibfnamefont
  {E.}~\bibnamefont {Greplova}},\ }\bibfield  {title} {\bibinfo {title}
  {Automated tuning of double quantum dots into specific charge states using
  neural networks},\ }\href {https://doi.org/10.1103/PhysRevApplied.13.054019}
  {\bibfield  {journal} {\bibinfo  {journal} {Phys. Rev. Appl.}\ }\textbf
  {\bibinfo {volume} {13}},\ \bibinfo {pages} {054019} (\bibinfo {year}
  {2020})}\BibitemShut {NoStop}%
%
\bibitem {Ziegler23-AEC}%
  \BibitemOpen
  \bibfield  {author} {\bibinfo {author} {\bibfnamefont {J.}~\bibnamefont
  {Ziegler}}, \bibinfo {author} {\bibfnamefont {F.}~\bibnamefont {Luthi}},
  \bibinfo {author} {\bibfnamefont {M.}~\bibnamefont {Ramsey}}, \bibinfo
  {author} {\bibfnamefont {F.}~\bibnamefont {Borjans}}, \bibinfo {author}
  {\bibfnamefont {G.}~\bibnamefont {Zheng}},\ and\ \bibinfo {author}
  {\bibfnamefont {J.~P.}\ \bibnamefont {Zwolak}},\ }\bibfield  {title}
  {\bibinfo {title} {Automated extraction of capacitive coupling for quantum
  dot systems},\ }\href {https://doi.org/10.1103/PhysRevApplied.19.054077}
  {\bibfield  {journal} {\bibinfo  {journal} {Phys. Rev. Appl.}\ }\textbf
  {\bibinfo {volume} {19}},\ \bibinfo {pages} {054077} (\bibinfo {year}
  {2023}{\natexlab{b}})}\BibitemShut {NoStop}%
%
\bibitem {Zwolak18-QLD}%
  \BibitemOpen
  \bibfield  {author} {\bibinfo {author} {\bibfnamefont {J.~P.}\ \bibnamefont
  {Zwolak}}, \bibinfo {author} {\bibfnamefont {S.~S.}\ \bibnamefont
  {Kalantre}}, \bibinfo {author} {\bibfnamefont {X.}~\bibnamefont {Wu}},
  \bibinfo {author} {\bibfnamefont {S.}~\bibnamefont {Ragole}},\ and\ \bibinfo
  {author} {\bibfnamefont {J.~M.}\ \bibnamefont {Taylor}},\ }\bibfield  {title}
  {\bibinfo {title} {{QFlow} lite dataset: {A} machine-learning approach to the
  charge states in quantum dot experiments},\ }\href
  {https://doi.org/10.1371/journal.pone.0205844} {\bibfield  {journal}
  {\bibinfo  {journal} {PLoS ONE}\ }\textbf {\bibinfo {volume} {13}},\ \bibinfo
  {pages} {e0205844} (\bibinfo {year} {2018})}\BibitemShut {NoStop}%
%
\bibitem {Fedele21-SOT}%
  \BibitemOpen
  \bibfield  {author} {\bibinfo {author} {\bibfnamefont {F.}~\bibnamefont
  {Fedele}}, \bibinfo {author} {\bibfnamefont {A.}~\bibnamefont {Chatterjee}},
  \bibinfo {author} {\bibfnamefont {S.}~\bibnamefont {Fallahi}}, \bibinfo
  {author} {\bibfnamefont {G.~C.}\ \bibnamefont {Gardner}}, \bibinfo {author}
  {\bibfnamefont {M.~J.}\ \bibnamefont {Manfra}},\ and\ \bibinfo {author}
  {\bibfnamefont {F.}~\bibnamefont {Kuemmeth}},\ }\bibfield  {title} {\bibinfo
  {title} {Simultaneous operations in a two-dimensional array of
  singlet-triplet qubits.},\ }\href
  {https://doi.org/10.1103/PRXQuantum.2.040306} {\bibfield  {journal} {\bibinfo
   {journal} {PRX Quantum}\ }\textbf {\bibinfo {volume} {2}},\ \bibinfo {pages}
  {040306} (\bibinfo {year} {2021})}\BibitemShut {NoStop}%
%
\bibitem {Sohn13}%
  \BibitemOpen
  \bibfield  {author} {\bibinfo {author} {\bibfnamefont {L.~L.}\ \bibnamefont
  {Sohn}}, \bibinfo {author} {\bibfnamefont {L.~P.}\ \bibnamefont
  {Kouwenhoven}},\ and\ \bibinfo {author} {\bibfnamefont {G.}~\bibnamefont
  {Sch{\"o}n}},\ }\href@noop {} {\emph {\bibinfo {title} {Mesoscopic electron
  transport}}}\ (\bibinfo  {publisher} {{Springer Netherlands}},\ \bibinfo
  {year} {1997})\BibitemShut {NoStop}%
%
\bibitem {Baart16-CAT}%
  \BibitemOpen
  \bibfield  {author} {\bibinfo {author} {\bibfnamefont {T.~A.}\ \bibnamefont
  {Baart}}, \bibinfo {author} {\bibfnamefont {P.~T.}\ \bibnamefont {Eendebak}},
  \bibinfo {author} {\bibfnamefont {C.}~\bibnamefont {Reichl}}, \bibinfo
  {author} {\bibfnamefont {W.}~\bibnamefont {Wegscheider}},\ and\ \bibinfo
  {author} {\bibfnamefont {L.~M.~K.}\ \bibnamefont {Vandersypen}},\ }\bibfield
  {title} {\bibinfo {title} {Computer-automated tuning of semiconductor double
  quantum dots into the single-electron regime},\ }\href
  {https://doi.org/10.1063/1.4952624} {\bibfield  {journal} {\bibinfo
  {journal} {Appl. Phys. Lett.}\ }\textbf {\bibinfo {volume} {108}},\ \bibinfo
  {pages} {213104} (\bibinfo {year} {2016})}\BibitemShut {NoStop}%
%
\bibitem {Note1}%
  \BibitemOpen
  \bibinfo {note} {We use a notation value(uncertainty) to express
  uncertainties, for example, $1.5(6)~\si {\centi \meter }$. All uncertainties
  herein reflect the uncorrelated combination of single-standard deviation
  statistical and systematic uncertainties.}\BibitemShut {Stop}%
%
\bibitem {Fedele19-PhD}%
  \BibitemOpen
  \bibfield  {author} {\bibinfo {author} {\bibfnamefont {F.}~\bibnamefont
  {Fedele}},\ }\emph {\bibinfo {title} {Spin interactions within a
  two-dimensional array of GaAs double dots}},\ \href
  {https://nbi.ku.dk/english/theses/phd-theses/federico-fedele/Federico_Fedele.pdf}
  {Ph.D. thesis},\ \bibinfo  {school} {Niels Bohr Institute, Faculty of
  Science, University of Copenhagen}, \bibinfo {address} {Copenhagen, Denmark}
  (\bibinfo {year} {2019})\BibitemShut {NoStop}%
\end{thebibliography}
\end{document}